\documentclass[prb,aps,showpacs,twocolumn,amssymb,amsmath,amstext,amsfont,noshowkeys]{revtex4}
\usepackage{graphicx}
\usepackage{theorem}
\usepackage{dcolumn}
\usepackage{bm}

\newcommand{\Dd}{{\rm d}}

\begin{document}

\title{Stability of multi-electron bubbles in liquid helium}
\author{Wei Guo}\author{Dafei Jin}\author{Humphrey J. Maris} \affiliation{Department of Physics, Brown University,
Providence, RI 02912, U.S.A.}
\date{\today}

\begin{abstract}

The stability of multi-electron bubbles in liquid helium is investigated theoretically. We find
that multi-electron bubbles are unstable against fission whenever the pressure is positive. It is
shown that for moving bubbles the Bernoulli effect can result in a range of pressures over which
the bubbles are stable.

\end{abstract}

\pacs{67.40.Yv, 47.55.Dz, 73.20.-r} \maketitle

\section{Introduction}

Multi-electron bubbles in liquid helium were first observed by Volodin \textit{et
al}.\cite{VolodinJETPL} In their experiment a layer of electrons was held in place just above the
free surface of a bath of liquid helium by an electric field. The field was produced by a positive
voltage applied to an electrode immersed in the liquid. The electrons remained outside the helium
because for an electron to enter liquid helium it has to overcome a potential barrier of height
approximately 1 eV.\cite{SommerPRL} When the field reached a critical value, the surface of the
liquid became unstable and a large number of electrons entered into the liquid through the
formation of bubbles. Each of these bubbles typically contained $10^7\sim10^8$ electrons. The
multi-electron bubbles are of interest because they could possibly provide a way to study a number
of properties of an electron gas on a curved surface.\cite{TempereSSR}

As a first approximation, one can consider that the radius of a spherical multi-electron bubble
(MEB) is such as to minimize the sum of the energy associated with the Coulomb repulsion of the
electrons and the surface energy of the bubble. This gives an equilibrium radius of
\begin{equation}
R_0=\left(\frac{Z^2e^2}{16\pi\sigma\epsilon}\right)^{1/3},
\end{equation}
where $Z$ is the number of electrons, $\sigma$ is the surface tension of helium (0.36 erg~cm$^{-2}$
at 1.3 K),\cite{VicentePRB} $\epsilon$ is the dielectric constant (1.0573 at low temperature), and
the applied pressure has for the moment been taken to be zero. Thus, for example, for $N=10^7$ the
radius is 106 $\mu$m.

So far, there have been a very limited number of experimental studies of these
bubbles.\cite{VolodinJETPL,KhaikinJPPC,AlbrechtEPL,AlbrechtJLTP} In this paper we first consider
the stability of an MEB that is at rest in the liquid (section II). We find that, at least when the
simplest model of the energy of the electron system is used, the bubble is unstable against fission
whenever the applied pressure is positive. In section III we investigate how the stability of a
bubble is changed when it is moving through the liquid. We have been able to determine the region
in the pressure-velocity plane where the bubble is stable.

\section{STABILITY OF BUBBLES AT REST}

Since MEB's were first observed, there have been several theoretical investigations of the
stability of these objects. The first discussion was given by Shikin\cite{ShikinJETPL} and further
analysis has been given by Salomaa and Williams,\cite{SalomaaPRL,SalomaaPS,HannahsIEEEUS}  and
Tempere, Silvera and coworkers.\cite{TempereSSR,TemperePRL,SilveraPSSB,TemperePRB} In the simplest
model, the electrons are taken to be distributed over the inner surface of the bubble in a way such
that the electric field is everywhere exactly normal to the surface. This ensures that the charge
distribution is in equilibrium. The electrons are treated classically and so are localized at the
surface in a layer of zero thickness (see below). Thus the total energy of the bubble is taken to
be
\begin{equation}
E=E_S+E_V+E_C.
\end{equation}
Here $E_S=\sigma S$ is the surface energy with $S$ the surface area and $\sigma$ the surface
tension, $E_V=PV$ the volume energy with $P$ the applied pressure and $V$ the bubble volume, and
$E_C$ is the Coulomb energy given by
\begin{equation}
E_C=\int\frac{\epsilon \mathcal{E}^2}{8\pi}\ \Dd V. \label{Coulomb}
\end{equation}
Since the electrons can move freely around the surface, the field $\mathcal{E}$ inside the bubble
must be zero and so the integral in Eq.~(\ref{Coulomb}) can be restricted to the region outside the
bubble. If the bubble is spherical, the bubble radius that gives the minimum value of the energy is
the solution of the equation
\begin{equation}
R_0=\left(\frac{Z^2e^2}{16\pi\epsilon\sigma+8\pi\epsilon
PR_0}\right)^{1/3}.
\end{equation}
For zero applied pressure, this gives the total energy of an MEB as
\begin{equation}
E=\frac{3}{2}\left(2\pi Z^4e^4\sigma\epsilon^2\right)^{1/3}.
\end{equation}
Since the energy is proportional to $Z^{4/3}$ the energy is always reduced if the bubble breaks
into two. Hence, in the discussion of stability given here we are not considering whether the
energy of the bubble can be lowered if it breaks into pieces, but are trying to determine whether
there is an energy barrier that prevents the bubble from breaking.

To consider whether the spherical shape is stable, write
\begin{equation}
R(\theta,\phi)=R_0\left\{1+\sum_{l=0}^{\infty}\sum_{m=-l}^{l}\eta_{lm}Y_{lm}(\theta,\phi)\right\},
\label{Expansion}
\end{equation}
where $\eta_{l,-m}=\eta^*_{lm}$. It is straightforward to show that to second order in the
parameters $\eta_{lm}$, the three contributions to the energy can be written as
\begin{equation}
\begin{split}
E_S=\ & 4\pi\sigma R_0^2
\left\{1+\frac{1}{\sqrt{\pi}}\eta_{00}\right.
\\
& + \left. \frac{1}{8\pi} \sum_{l=0}^{\infty}\sum_{m=-l}^{l} (l^2+l+2) |\eta_{lm}|^2\right\},
\end{split}
\end{equation}
\begin{equation}
\begin{split}
E_V=\ & \frac{4\pi}{3} PR_0^3
\left\{1+\frac{3}{2\sqrt{\pi}}\eta_{00}\right.
\\
& + \left. \frac{3}{4\pi}\sum_{l=0}^{\infty}\sum_{m=-l}^{l}|\eta_{lm}|^2\right\},
\end{split}
\end{equation}
\begin{equation}
\begin{split}
E_C=\ & \frac{Z^2e^2}{2\epsilon R_0} \left\{1-\frac{1}{2\sqrt{\pi}}\eta_{00} +
\frac{1}{4\pi}\eta_{00}^2 \right.
\\
& - \left. \frac{1}{4\pi}\sum_{l=0}^{\infty}\sum_{m=-l}^{l}l|\eta_{lm}|^2\right\}.
\end{split}
\end{equation}
Hence the total energy is
\begin{equation}
E=12\pi\sigma R_0^2+\frac{16\pi}{3}PR_0^3 + \sum_{l=0}^{\infty}
\frac{1}{2}\alpha_l \sum_{m=-l}^{l} |\eta_{lm}|^2,
\end{equation}
where the spring coefficients $\alpha_l$ are given by
\begin{equation}
\alpha_0=3\sigma R_0^2+2PR_0^3,
\end{equation}
and
\begin{equation}
\alpha_l=(l-2)(l-1)\sigma R_0^2-2(l-1)PR_0^3,\label{SpringI}
\end{equation}
for $l\geq 1$. From this one can see that the bubble is stable against spherically symmetric
perturbations provided that $3\sigma R_0^2+2PR_0^3>0$. This leads to the condition $P>P_c$ where
\begin{equation}
P_c=-\left(\frac{27\pi\epsilon\sigma^4}{2Z^2e^2}\right)^{1/3}.
\end{equation}
For $l =1$, the spring coefficient $\alpha_1$ is zero; this is to be expected since a perturbation
of the form $\eta_{1m}Y_{1m}(\theta,\phi)$ corresponds to a simple translation of the bubble in
some direction. For $l=2$ the spring constant $\alpha_2$ is zero if the pressure is zero, and so
this analysis of the effect of small perturbations to the initial spherical shape does not
determine the stability of the bubble. The higher $l$ spring constants are all positive at zero
pressure but each becomes negative if the pressure is increased to a sufficiently positive value.
It was noted by Tempere \textit{et al}.\cite{TemperePRL} that if the pressure is negative (but not
negative with respect to $P_c$), all of the spring constants will be positive\cite{MorePrecise} and
so the bubble must be stable.

The stability of the bubble at zero pressure is of especial importance since in the experiments
that have been performed so far there has been no applied pressure apart from the very small
hydrostatic pressure due to the distance the bubble is below the free surface. At zero pressure
$\alpha_2$ is zero, and so we need to go beyond the lowest order in perturbation theory in order to
investigate the stability of an MEB at zero pressure. One approach would be to calculate the terms
in the energy that are fourth order in the $\eta_{lm}$ parameters. Instead we have performed
numerical calculations of the total energy as a function of bubble shape.

To do this, we describe the shape of the bubble using Eq.~(\ref{Expansion}) but now do not restrict
the parameters $\eta_{lm}$ to being small. When the bubble shape changes, the electrons will
redistribute themselves over the surface so as to minimize the energy and to make the electric
field inside the bubble zero. For each choice of shape we use the finite element
method\cite{JinWiley} to compute the surface charge distribution and the Coulomb energy. The
simulation uses 1280 triangle patches. We start with a spherical shape and vary the parameters
$\eta_{lm}$ to see if a state of lower energy can be reached without passing over a barrier. We
have done this using a maximum value of 5 for $l$ in Eq.~(\ref{Expansion}). This process was then
repeated for a series of different pressures. We also performed similar calculations with a maximum
value of $l$ of 15 but taking only $m=0$. Both procedures gave the same results for the stability.

\begin{figure*}
\includegraphics[scale=0.9]{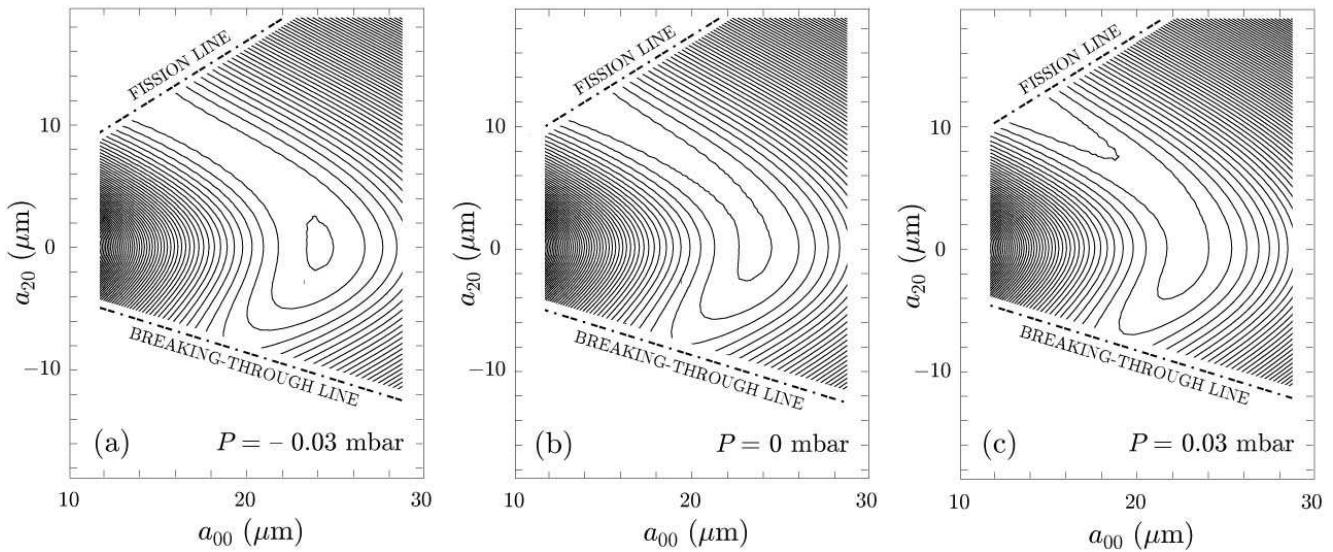}
\caption{Contour lines of constant energy for a multi-electron bubble containing $10^6$ electrons
for three different pressures. The energy spacing between contour lines is 0.05 eV. The energy is
shown as a function of the parameters $a_0$ and $a_2$ as defined in Eq.~(\ref{SimpleExpansion}).
The electrons are distributed over the bubble surface so as to minimize their
energy.}\label{ContourI}
\end{figure*}

The result of this investigation is that for all positive pressures there is no barrier to fission,
whereas for negative pressures there is a barrier. This result holds for all values of $Z$. To
illustrate the path to fission, we describe results obtained for a simplified calculation in which
only $l=0$ and $l=2$ contributions are retained. Thus we write
\begin{equation}
R(\theta,\phi)=a_0+a_2(3\cos^2\theta-1).\label{SimpleExpansion}
\end{equation}
Within this simplified model, fission occurs when $a_2=a_0$ and the bubble develops a hole along
the $z$-axis, i.e., takes on a donut shape, when $a_2=-a_0/3$.  In Fig.~\ref{ContourI}(a), we show
examples of contour plots of the energy in the $a_0$-$a_2$ plane. The pressure is $-0.03$ mbar and
$Z=10^6$. There is a stable minimum with $a_2$ equal to zero, i.e., the bubble is spherical. When
the pressure is zero (Fig.~\ref{ContourI}(b)), there is still a point in the plane at which the
energy of the bubble is stationary with respect to both $a_0$ and $a_2$ (at $a_0=23.8\ \mu$m and
$a_2=0$), but it is now possible to reach the fission line from this point without passing over any
energy barrier. Note that along this path there is, of course, an increase in the value of $a_2$
but also a substantial decrease in $a_0$. Once the pressure becomes positive (see, for example,
Fig.~\ref{ContourI}(c)), there is no point in the $a_0$-$a_2$ plane where the energy is stationary.

These results can be compared with the earlier calculations by Tempere \textit{et
al}.\cite{TemperePRB} who also investigated the stability against fission. They used an ingenious
method in which the bubble was described by 6 parameters chosen so that the shape of a bubble
undergoing fission could consist of two spheroids connected by a hyperboloidal neck. The choice of
parameters was such that the bubble could vary from consisting of a single sphere, to an ellipsoid,
and then all the way to separated spheres. They minimized the total energy of the bubble by
adjusting these parameters subject only to the constraint that the total length $L$ of the bubble
had to have a given value. They then investigated how the total energy varied with $L$ starting
from a value of $L$ equal to $2R_0$. If the energy decreased monotonically as $L$ increased from
$2R_0$ to a large value, this indicated that the MEB was unstable against fission. If the energy
first increases before decreasing, this indicates that the bubble is stable. To simplify the
calculation, Tempere \textit{et al.} made the approximation that the charge density was uniform
over the surface of the bubble. They concluded that at zero pressure even though there is a mode of
deformation (the $l=2$ mode) which can grow without increasing the energy of the bubble, there
should be an energy barrier which prevents fission,\cite{SeeConclusion} whereas we find no barrier.

\begin{figure}
\includegraphics[scale=0.3]{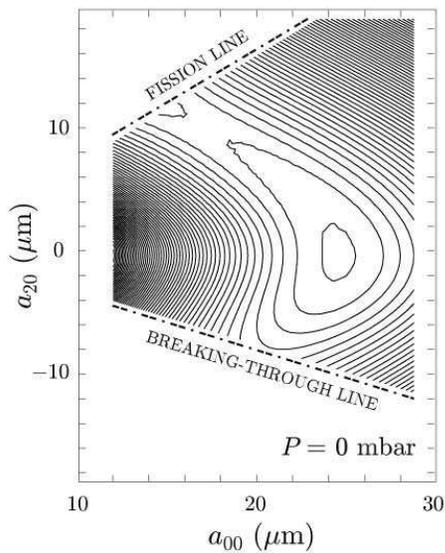}
\caption{Contour lines of constant energy for a multi-electron bubble containing $10^6$ electrons
at zero pressure. The energy spacing between contour lines is 0.05 eV.  The energy is shown as a
function of the parameters $a_0$ and $a_2$ as defined in Eq.~(\ref{SimpleExpansion}). The electrons
are uniformly distributed over the surface area of the bubble.}\label{ContourII}
\end{figure}

This difference in the results arises from the treatment of the charge distribution on the bubble.
If the bubble is assumed to have surface charge density that remains uniform when the shape
changes, it is straightforward to show that the Coulomb energy for small changes from the
equilibrium spherical shape is
\begin{equation}
\begin{split}
E_C=\ & \frac{Z^2e^2}{2\epsilon R_0} \left\{1-\frac{1}{2\sqrt{\pi}}\eta_{00} +
\frac{1}{4\pi}\eta_{00}^2 \right.
\\
& - \left.
\frac{1}{4\pi}\sum_{l=0}^{\infty}\sum_{m=-l}^{l}\frac{l^2+3l-1}{2l+1}|\eta_{lm}|^2\right\}.
\end{split}
\end{equation}
In this case the spring constant $\alpha_l'$ for the $l$-th mode (considering only $l\geq1$)
becomes
\begin{equation}
\alpha_l'=(l^2+l+2)\sigma R_0^2+2PR_0^3-\frac{l^2+3l-1}{2l+1}\frac{Z^2e^2}{4\pi\epsilon R_0}.
\end{equation}
Comparing this with the spring constant $\alpha_l$ when the charge redistributes
(Eq.~(\ref{SpringI})) gives
\begin{equation}
\alpha_l'=\alpha_l+\frac{(l-1)^2}{2l+1}\frac{Z^2e^2}{4\pi\epsilon R_0}.
\end{equation}
Thus for all modes, except $l=0$ and $l=1$, making the approximation of a uniform surface charge
gives an increase in stiffness and makes it harder for the bubble to undergo fission. The increase
in stiffness is to be expected since a redistribution of surface charge can only lower the total
energy. In Fig.~\ref{ContourII}, we show energy contour lines in the $a_0$-$a_2$ plane for an MEB
with $10^6$ electrons at zero pressure calculated by taking a uniform surface charge. One can see
that within this approximation the spherical bubble is stable.

There are several physical effects that are not included in the simplified model used so far. It is
possible that allowance for these effects would change the stability of an MEB at zero pressure. A
more detailed consideration of the Coulomb energy (the total electron energy, to be more precise)
for a spherical bubble has been given by Salomaa and Williams\cite{SalomaaPRL,SalomaaPS} using the
density functional formalism of Hohenberg and Kohn.\cite{HohenbergPR} This makes possible the
inclusion of the kinetic, exchange and correlation energies, but how these extra contributions
affect the spring constants is not clear and is difficult to calculate. Salomaa and Williams show
that these extra contributions to the energy make a very small contribution to the energy when $Z$
is large. For example, for $Z=10^8$ the extra terms make a contribution that is roughly 4000 times
smaller than the form $Z^2e^2/2\epsilon R_0$ for the energy used in the simple model. The
calculation could also be improved, for example, by using a density functional theory to treat the
surface of the liquid helium, and by allowing for the penetration of the electron wave function
into the liquid. All of these effects appear to be very small corrections to the total energy and
hence are unlikely to change the spring constants by a large amount. However, it is important to
note that even a small correction could lead to a positive value for $\alpha_{2}$ which would in
turn lead to a finite (but small) energy barrier against fission. As an example, consider
corrections that arise as a result of using a density-functional scheme to describe the helium. For
a bubble with radius large compared to the thickness of the liquid-vapor interface the first
correction to the energy can be represented by considering the surface tension $\sigma$ to contain
a correction that is proportional to the total curvature $\kappa$ of the surface. Based on a simple
density functional scheme used previously,\cite{XiongJLTP} it is straightforward to show that the
correction to the surface tension is $\Delta\sigma=\sigma'\kappa$, where
\begin{equation}
\sigma'=0.9\times10^{-8}  \mathrm{\ erg\ cm}^{-1},
\end{equation}
and the sign of the correction is such that the surface tension is increased for a concave surface
of the liquid. Inclusion of this term changes the total energy by an amount $\Delta E$ which for a
bubble at zero pressure is given by
\begin{equation}
\frac{\Delta E}{E}=\frac{0.08}{Z^{2/3}}.
\end{equation}
It is straightforward to show that the spring constant for an $l=2$ deformation at zero pressure
now becomes
\begin{equation}
\alpha_2=12\sigma'R_0.
\end{equation}
Because $\alpha_2$ is now positive at zero pressure there will be a barrier against fission, but
clearly for large $Z$ (e.g. $Z\sim10^8$), this barrier will be very small.

\section{STABILITY OF MOVING BUBBLES}

The above results indicate that one way to stabilize an MEB is to produce it in liquid that is
under a small negative pressure. We now consider an alternate way to maintain a stable bubble. A
bubble moving through a liquid will be affected by the local pressure change associated with the
liquid moving around it. For a spherical bubble moving at velocity $v$ through an incompressible
inviscid fluid with density $\rho$, the Bernoulli effect results in a pressure variation over the
surface of the bubble which is given by\cite{LandauPergamon}
\begin{equation}
\begin{split}
P(\theta) &=P_0+\frac{1}{8}\rho v^2(9\cos^2\theta-5)\\
&=P_0-\rho v^2\sqrt{\frac{\pi}{2}}Y_{00}(\theta,\phi)+\rho
v^2\sqrt{\frac{9\pi}{20}}Y_{20}(\theta,\phi).
\end{split}
\end{equation}
For a bubble in liquid that is at zero pressure far removed from the bubble ($P_0=0$), this changes
the shape of the bubble in two ways. The term proportional to $Y_{00}$ by itself would provide a
negative pressure around the surface of the bubble and since bubbles are stable at negative
pressure, this contribution serves to stabilize the bubble. The second term gives a positive
pressure at the poles of the bubble and a negative pressure around the waist. This pressure
distribution will distort a spherical bubble so as to make the parameter $\eta_{20}$ in
Eq.~(\ref{Expansion}), or $a_2$ in Eq.~(\ref{SimpleExpansion}), to be negative. This tends to
stabilize the bubble since, as can be seen from Fig.~\ref{ContourI}, for fission to occur $a_2$ has
to become positive.

\begin{figure}
\includegraphics[scale=0.3]{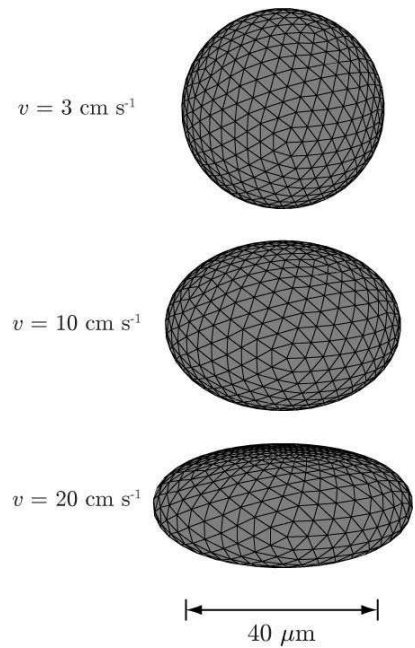}
\caption{The shape of a multi-electron bubble containing $10^6$ electrons for bubble velocities of
3, 10 and 20 cm~s$^{-1}$.}\label{Shape}
\end{figure}

\begin{figure}
\includegraphics[scale=0.42]{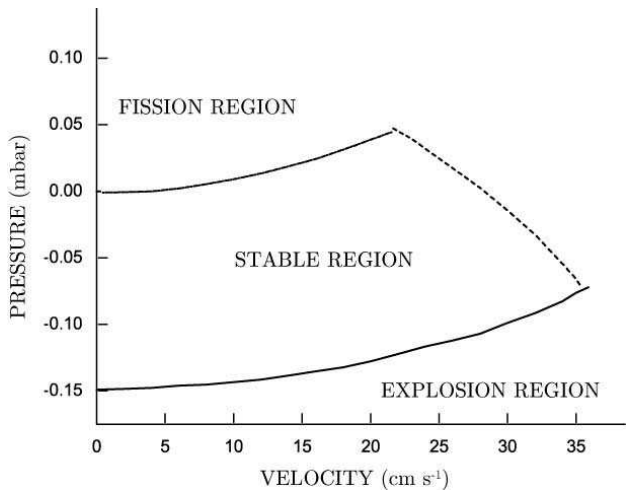}
\caption{Plot of the region in the pressure-velocity plane in which a MEB containing $10^6$
electrons is stable. The region is bounded by the lines on which the two different types of
instability occur as described in the text. Along the dashed line the bubble becomes concave at the
poles and the numerical calculations become inaccurate.}\label{Phase}
\end{figure}

We have performed computer simulations in order to find the shape of moving bubbles and the range
of velocity and pressure for which they are stable. We start with a guess at the bubble shape and
then calculate the charge distribution on the surface. This then gives the pressure $\Delta
P_{\mathrm{el}}(\theta)$ exerted on the surface by the electrons. We then find the flow in the
liquid. To do this we expand the velocity potential as
\begin{equation}
\Phi(\theta)=\sum_l B_lP_l(\cos\theta)r^{-l-1},
\end{equation}
where $B_l$ are some coefficients and the sum includes terms from $l=1$ to $l=20$. The coefficients
are determined so as to give a velocity distribution in the liquid such that in the frame of
reference of the moving bubble, the liquid velocity at the bubble surface in the direction normal
to bubble surface is as close to zero as possible. This gives a pressure at the bubble surface of
\begin{equation}
P_0+\Delta P_{\mathrm{B}}(\theta),
\end{equation}
where $P_0$ is the pressure in the bulk liquid far removed from the bubble and $\Delta
P_{\mathrm{B}}(\theta)$ is the Bernoulli pressure. The net inward force acting on unit area of the
bubble surface is then
\begin{equation}
P_0+\Delta P_{\mathrm{B}}(\theta)+2\sigma\kappa-\Delta P_{\mathrm{el}}(\theta),
\end{equation}
where $\kappa$ is the total curvature of the surface and $\Delta P_{\mathrm{el}}(\theta)$ is the
outward pressure exerted by the electrons. Each part of the bubble surface is then moved inward a
distance proportional to this force, and the process repeated until the equilibrium shape is found.
The calculation used a maximum value of $l$ of 15. The calculation as just described is based on
the assumption that the bubble shape and velocity field have axial symmetry around the direction in
which the bubble is moving. In order to test this assumption, we also performed a calculation in
which axial symmetry was not assumed. This calculation used values of $l$ up to 5 and $|m|\leq5$.
This calculation showed that the axially symmetric shape was stable. For an MEB with $Z=10^6$
shapes for three velocities are shown in Fig.~\ref{Shape}. We are able to perform the numerical
calculation until the bubble becomes concave at the poles. This is shown by the dashed line in
Fig.~\ref{Phase}.

\begin{figure}
\includegraphics[scale=0.38]{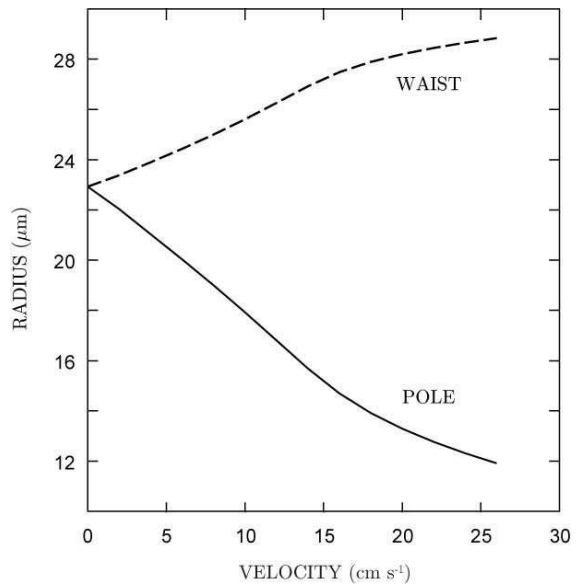}
\caption{Distance $R_{\mathrm{pole}}$ from the bubble center to the
poles and the radius $R_{\mathrm{waist}}$ of the waist as a function
of the bubble velocity. These results are for a bubble containing
$10^6$ electrons moving through liquid in which the pressure at
large distance from the bubble is zero.}\label{ShapeParameter}
\end{figure}

In Fig.~\ref{ShapeParameter}, the distance $R_{\mathrm{pole}}$ from the bubble center to the pole
and the radius $R_{\mathrm{waist}}$ of the waist are shown as a function of the velocity. In
Fig.~\ref{Phase} we show a plot of the region in the pressure-velocity plane in which the bubble is
stable. This region is bounded by two lines. For small velocities there is a critical positive
pressure at which the bubble undergoes fission. At negative pressures the bubble becomes unstable
against expansion. For zero velocity this expansion is isotropic.

Note that the change in the shape of the bubble even for a small velocity is surprisingly large.
This comes about simply because the Bernoulli pressure contains a finite term varying with angle as
$Y_{20}(\theta,\phi)$ but the spring constant $\alpha_2$ for this pressure component is zero. Thus,
for an MEB the changes in $R_{\mathrm{pole}}$ and $R_{\mathrm{waist}}$ are linearly proportional to
the bubble velocity whereas for a gas bubble in a liquid the spring constant $\alpha_2$ is finite
and so the changes in dimensions are proportional to the square of the velocity.

The region of stability of bubbles containing a different number of electrons can be found by
scaling the results shown in Fig.~\ref{Phase}. The instability pressure $P_{\mathrm{instab}}(Z,v)$
can be written in the form
\begin{equation}
P_{\mathrm{instab}} (Z,v)= A Z^{-2/3}\left(\frac{\sigma^4\epsilon}{Z^2e^2}\right)^{1/3}
f(Bv^2Z^{2/3}),
\end{equation}
where $f$ is a dimensionless function,
$A=(\sigma^4\epsilon/e^2)^{1/3}$ and
$B=\rho(e^2/\sigma^4\epsilon)^{1/3}$. Thus, for zero velocity the
critical negative pressure at which a bubble becomes unstable is
proportional to $Z^{-2/3}$, and  at zero applied pressure the
critical velocity at which the bubble becomes concave at the poles
occurs is proportional to $Z^{-1/3}$.

We note that in this paper we have treated the liquid as inviscid although, of course, helium above
the lambda point has a finite viscosity and below the lambda point the liquid still has a normal
fluid component. At sufficiently low temperatures the density of the normal fluid becomes very
small and, in addition, the mean free path of the excitations making up the normal fluid becomes
comparable to the radius of an MEB. Under these conditions, it appears that the only effect of the
normal fluid is to determine the mobility of an MEB and there should be no effect on the shape
change or the stability. For a bubble with $Z=10^6$ the mean free path of the thermal excitations
becomes equal to the radius at around 0.6 K and at this temperature the normal fluid density is
less than the total density by a factor of $4\times10^{-5}$. But as far as we are aware, there have
been no experiments with MEB's at low temperature.

At high temperatures where the helium is in the normal state, the situation is not so clear. It is
known that when the Reynolds number is large (but not so large that the flow becomes turbulent) the
viscosity results in a thin boundary layer on the surface of the bubble and the pressure at the
bubble surface is close to the value that would result from potential flow.\cite{SeeExample} This
general idea would suggest that the inviscid approximation should give reliable results for the
stability of MEB's over a wide range of Reynolds number. To determine this range one could
calculate the effect of viscosity using the method developed by Li and Yan\cite{LiNHT} and applied
by them to calculate the shape and drag on gas bubbles moving through a liquid. We have not
attempted to do this. We note that, Albrecht and Leiderer\cite{AlbrechtJLTP} in their experiments
at 3.5 K found that the mobility of the MEB's was between one and two orders of magnitude larger
than expected on the basis of ordinary hydrodynamics. The reason for this is unknown.

In the experiments of Volodin \textit{et al.}\cite{VolodinJETPL} and Khaikin\cite{KhaikinJPPC}
which were performed at 1.3 K, velocities of the order of $10^4$ cm~s$^{-1}$ were reported. At
these velocities the normal fluid component would be in the turbulent regime and the bubble is
moving so fast that it should lose energy through the production of quantized vortex rings.

\section{Conclusion}

We have examined the stability of multi-electron bubbles in liquid helium and found that stationary
bubbles at positive pressures are unstable. We show that because of the Bernoulli effect moving
bubbles can be stable even at small positive pressures.

\begin{acknowledgments}

This work was supported by the National Science Foundation through Grant No. DMR-0605355.

\end{acknowledgments}


\begin{thebibliography}{99}
\bibitem{MorePrecise} To be more precise, we should say all except the spring for $l=1$  which is always zero.
\end{thebibliography}
\end{document}